# Metalenses at visible wavelengths: past, present, perspectives

Authors: Philippe Lalanne[1]*, Pierre Chavel[2,3]*

[1] Laboratoire Photonique, Numérique et Nanosciences (LP2N), IOGS – Univ. Bordeaux – CNRS, 33400 Talence cedex, France.

[2] Laboratoire Charles Fabry (LCF), IOGS – CNRS – Univ. Paris-Saclay, 2 avenue Augustin Fresnel, 91127 Palaiseau cedex, France.

[3] Laboratoire Hubert Curien, Université de Lyon, Univ. Jean Monnet de Saint Etienne, CNRS UMR 5516, 42000 Saint Etienne, France

*Correspondence to: philippe.lalanne@institutoptique.fr, pierre.chavel@institutoptique.fr

**Abstract**: The so-called 'flat optics' that shape the amplitude and phase of light with high spatial resolution are presently receiving considerable attention. Numerous publications in high-impact journals seemingly offer hope for great promises for ultra-flat metalenses with high efficiency, high numerical aperture, broadband operation... We temperate the expectation by referring the current status of metalenses against their historical background, assessing the technical and scientific challenges recently solved and critically identifying those that still stand in the way.

## 1. Introduction

Recently, nanostructured metasurfaces have received increasing attention due to their ability to control the amplitude, phase, polarization, orbital momentum, absorption, reflectance, emissivity of light with high spatial resolution. This leads scientists and engineers to revisit various traditional applications in optics from a different angle. The broad panel of researches on metasurfaces, already well documented in excellent review articles [Zhu16, Kuz16], is not the topic of this expert opinion. We focus on metalenses hereafter, and since low numerical aperture (NA) lenses have been fabricated at low cost with remarkable performance for a long time, we restrict the discussion to high-NA, high efficiencies metalenses.

Since the publication of a research article in Science that revisited Snell's law at the interface between two uniform media thanks to an ultrathin grating composed of metallic nanoantennas etched on the interface [Yu11], the field of metalenses has been booming. A recent flagship research article [Kho16] that reports metalenses manufactured with a high NA operating at visible wavelengths is particularly emblematic of the growing interest. Comparisons with a state-of-the-art commercial objectives suggest that the image quality is as good and even better, and considering their flat nature and compact size, metalenses appear as potentially revolutionary. This new potential is all the more significant because classical échelette diffractive lenses present a ridiculously small efficiency for large NA [Swa89], precluding their use for high resolution focusing, except for the ancillary function of chromatic aberration correction where the local spatial frequency is typically much smaller.

However, metalenses do not come out of the blue and the so-called 'flat optics' that shapes the phase of free-space waves through *subwavelength structures* have a venerable history over many years from the microwave domain down to the visible. What is the paradigm change at the source of this anticipated revolution? New fabrication processes, new concepts, new applications, fundamental limitation shifts?

To answer these questions, we confront recent metalens achievements with an historical perspective of flat optical elements using subwavelength structures, to further analyze the fundamental limitations that have been lifted and better anticipate perspectives offered by flat subwavelength optics. This is precisely the outline of the present article, which starts with a brief overview of the fundamentals of flat optics and emphasizes approaches that offer both high numerical aperture and high diffraction efficiency, and potentially broadband operation.

## 2. Flat-optic basis

The basic role of optical components such as lenses or prisms is to shape incoming wavefronts. In recent years, various flat-optics components with an optical thickness on the order of the wavelength have been introduced and demonstrated. In contrast to bulk optics, where phase is modulated by a continuous variation of the optical thickness, flat optics restrict the modulation to its minimum, $2\pi$, by wrapping the bulk-optics phase modulo $2\pi$ for a nominal design wavelength $\lambda$. The wavefront is divided into zones wherein the phase varies by $2\pi$, with a $2\pi$ discontinuity between them. At $\lambda$, the diffraction patterns of the various zones combine coherently to form exactly the initial continuous wave, aside from border effects that are negligible if the zone widths are much larger than $\lambda$. At other wavelengths, the wrapping procedure introduces a specific dispersion effect that splits the wavefront into a set of orders and therefore reduces the diffraction efficiency of the desired order [Swa89]. Compared to travelling the same distance in air (of index assumed to be unity), the optical path delay for a zone thickness $t$ is

$$\Delta = (n-1)t. \qquad [1]$$

Conventional diffractive optical components [Aur72], such as the well-known "échelette" blazed grating, implement a $2\pi$-phase delay across the component surface by a $\lambda/(n-1)$ thickness excursion of the diffractive surface profile, see Fig. 1.

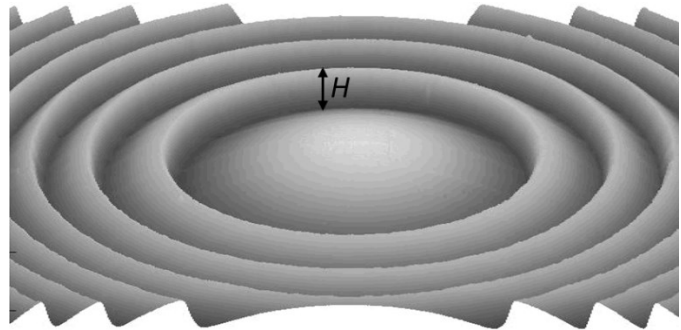

**Fig. 1. Echelette-type diffractive optical elements blazed in the first order.** Echelette lenses are often called Fresnel lenses, although the historical lighthouse Fresnel lens is essentially a bulk optical component. For a nominal wavelength $\lambda$, the Fresnel lens is said to be blazed into the $m^{\text{th}}$ order if the small facet thickness $H$ of every échelette is $m\lambda/(n-1)$, with $n$ being the refractive index of the lens material.

Nowadays, échelette-type diffractive optical elements are manufactured at low-cost by replication technologies such as embossing, molding and casting. Those technologies offer very high resolution, typically in the nanometer range, and allow the fabrication of large area, complex microstructures with minimal light loss in durable materials through high volume industrial production processes [Gal97]. However, they suffer two main limitations:

- The discontinuities of the wrapped phase introduce a shadow that wastes light into spurious orders. The effect is stringent for short period échelettes, to an extent that the échelette approach is essentially useless for high-NA lenses.

- The wrapping is valid a nominal design wavelength $\lambda$, implying that the imprinted phase varies with the wavelength of the incident light. Thus, the efficiency decreases as the illumination wavelength departs from the nominal one.

These fundamental limitations prevent a massive market rollout of échelette-Fresnel lenses and other diffractive optical elements in applicative areas. Indeed, imprinting the wrapped phase on a surface can be performed with structures other than the échelette. The art of flat-lens design is to devise recipes to lift the fundamental limitations of échelettes, minimizing the spurious effects introduced by the surface discontinuities. As will be shown, two generic approaches have been proposed. In the first approach, the phase results from a propagation delay (like for échelette), but implemented through an effective-index modulation and a waveguiding effect experienced by the transmitted wavefront. In the second approach, the phase is monitored by the scattering of nanoantennas with graded sizes or orientations. Both require a fine spatial sampling with subwavelength structures to faithfully imprint the rapid spatial variation of the wrapped phase at the outer zones of the lens.

## 3. High-NA metalens: an historical fresco

Because it combines both approaches (waveguiding and antenna emission), it is appropriate to start with the delay lens studied during the second world war by Kock [Koc48], who fabricated microwave diffractive lenses in high effective-index artificial dielectrics obtained by doping polystyrene foam sheets with subwavelength metallic insets, i.e. antennas. Drastic weight reduction was achieved with half-wavelength resonant insets, but broader frequency operation was achieved with even slightly smaller off-resonance insets. Amusingly, the delay results from both propagation and resonant scattering. The microwave domain also offers devices very similar, at least conceptually, to the metalenses that we will introduce hereafter, a particularly stringent case being the reflectarray antenna that utilizes arrays of waveguides to produce various beams [Ber63].

Extension to shorter wavelengths, did not occur until much more recently. Researchers in Erlangen [Sto91] and at MIT [Far92] were the first to propose monitoring the phase of graded-index artificial materials in the infrared, and then in the visible domain by a (relatively) slow local change of the profile of highly subwavelength gratings. The first experimental demonstrations of reflection gratings and a cylindrical mirror were performed at $\lambda = 10.6$ µm with metallic grooves of graded widths [Kip94].

Initial demonstrations of graded effective-index artificial-dielectrics manufactured for visible frequencies by etching quartz, glass and polymer films were disappointing, with measured diffraction efficiencies smaller than those of échelette components [Che95,Zho95,Mil96,Che96,War96]. At that time, the design mostly assumed an "adiabatic" effective index gradient, the analogy between local subwavelength gratings and artificial dielectrics was not properly understood, and modeling was challenging. Importantly too, the combined requirements of subwavelength spacing and $2\pi$-phase excursion led to aspect ratios difficult to manufacture with the materials and patterning technologies implemented.

New perspectives opened up by manufacturing graded artificial dielectrics components in high-index materials. For operation in the visible, $TiO_2$ appeared to be the most suitable transparent material. Figure 2 summarizes the main results in [Lal98,Lal99], which were obtained for a series of gratings with increasing deviation angle for operation at the $\lambda = 633$ nm HeNe laser line. The smallest pillar width is 90 nm, and the pillar spacing used to sample the phase is 272-nm. The gratings, called "blazed binary" components by the authors since they introduce a new type of blazed components derived from binary masking, were fabricated by direct e-beam writing, liftoff and chemically-assisted reactive ion etching using a process detailed in [Lal98]. Remarkably, efficiencies surpassing the hegemonic échelette components were observed for highly dispersive gratings corresponding to large deviation angles. The same concept applies to lenses: for the first time, one could therefore envision implementing efficient lenses with high-NA, removing the first limitation of échelette diffractive lenses.

Indeed, that demonstration was immediately followed by the fabrication of a 20° off-axis lens operating at $\lambda = 850$ nm with a vertical-cavity-surface-emitting laser for possible application to optical interconnects, see the SEM picture shown in Fig. 3c. The lens had an NA of 0.25 but since the it was designed for a 20° off-axis operation, the Fresnel zone widths were all smaller than $9\lambda$. Half of the aperture was composed of zones with widths smaller than $3\lambda$, corresponding to a much larger NA for an on-axis lens. Even so, the measured efficiency, including Fresnel reflections, reaches an impressive 80% [Lal99].

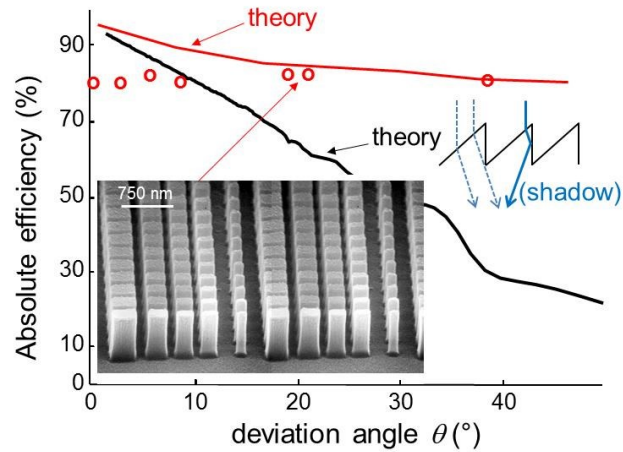

**Fig. 2. Blazed binary gratings beat their échelette counterparts with respect to the efficiency.** First-order transmission efficiencies of blazed-binary gratings as a function of the deviation angle $\theta$ in air (the grating period is varying) for a normally-incident and unpolarized plane wave at $\lambda = 633$ nm. Circles: experimental results. Red curve: theoretical results. The discrepancy at small $\theta$ is due to over etching fabrication errors resulting in a bond failure of tiny pillars. Black curve: theoretical results for échelette gratings in glass. The lower inset is an electron micrograph of a section of the grating operating at $\theta = 20°$; the period is 1.9 μm, the pillar aspect ratio reaches 4.6. The right inset shows the shadow zone responsible for efficiency drop of échelette gratings at large $\theta$'s. After [Lal98,Lal99,Lee02].

It is appropriate to underline why the échelette efficiencies were surpassed. For large deviation angles, the prism angle in each échelette period is as large as to cast shadow on the neighbors, wasting light into spurious orders, as illustrated in the right-hand side inset of Fig. 2 for normal incidence. The effect worsens at large deviation angle and at oblique incidence, as the relative impact of the shadow zone on the diffraction increases. In contrast, for high-index blazed-binary diffractive elements, every pillar behaves as an independent monomode waveguide that confines the light throughout the structure. No shadowing effect occurs even for oblique incidence [Lee02] with the small diffraction zones of high-NA components. In fact, such components operate at the limit of validity of the effective-index concept and behave analogously to phased arrays in radars [La99a]. Among other recent implementations, the most remarkable application of such components to date might be the Gaia grating [Zei10] sent to space in 2013 (period 3.15 µm, efficiency >80% over a 20.5 x 15.5 cm area).

The subject of metalenses has received considerably renewed attention recently with the introduction of resonance effects for monitoring the phase. In 2011, [Ish11] revived the interest of modulating the phase using metalic nanostructures to control the phase using the plasmonic dispersion inside a waveguiding slit in a metal and demonstrated focusing. Because resonance amplifies the phase delay compared to the propagation delay of Eq. 1, the requirement for large aspect ratios can be considerably relaxed. Indeed, the possibility to shape beams with graded resonant structures was considered in [Yu11], with plasmonic-based metasurfaces composed of tiny metallic nanoantennas. Although the research article insists on the deep-subwavelength thickness of the metasurface, see the Table in Fig. 3, the most important novelty is likely to be the demonstration that graded phases can be implemented by carefully designed nanoantennas. Following this finding, the domain has evolved quite rapidly. First it was realized that, in addition to Ohmic losses in the metal, single layer ultra-thin transmit arrays are reluctant to achieve $2\pi$-phase delays in co-polarization or impedance matching in cross polarization [Dat01,Arb14] and are thus efficient in reflection mode operation only [Hua13,Che14], and that it is advantageous, for transmission operation, to replace the nanoantennas by high-index dielectric subwavelength microresonators. Whence the recent revival of interest for dielectrics metalenses. In fact, because those components are thin, a slight absorption can be tolerated for some applications, so that silicon can be considered for use in the visible, as has been shown recently [Fat10,Vo14,Lin14,Wes14].

Overall, two physical effects are nowadays exploited to achieve the required $2\pi$-phase excursion, either a propagation delay to implement waveguide-type metalenses [Lal99,Zie10]) or a resonance delay [Vo14,Arb15,Yu15,Dec15] to implement resonant high-contrast metalenses by combining two resonances, each covering a standard phase range of $\pi$. Additionally, each effect may be implemented with centro-symmetric nanostructures to fabricate polarization insensitive metalenses or with rotationally-asymmetric nanostructures showing form birefringence. By rotating the nanostructures, full wavefront control can be achieved with a Berry-phase vortex [Has03] and a fancy mutual polarization behavior for right-handed and left-handed input-output beams arises [Lin14, Kho16].

By way of summary, Fig. 3 provides an historical overview of the concepts used so far for designing and fabricating flat optical wavefront shapers based on subwavelength binary nanostructures. One fairly exhaustive review of publications in the field appeared while this paper was being revised [Gen17].

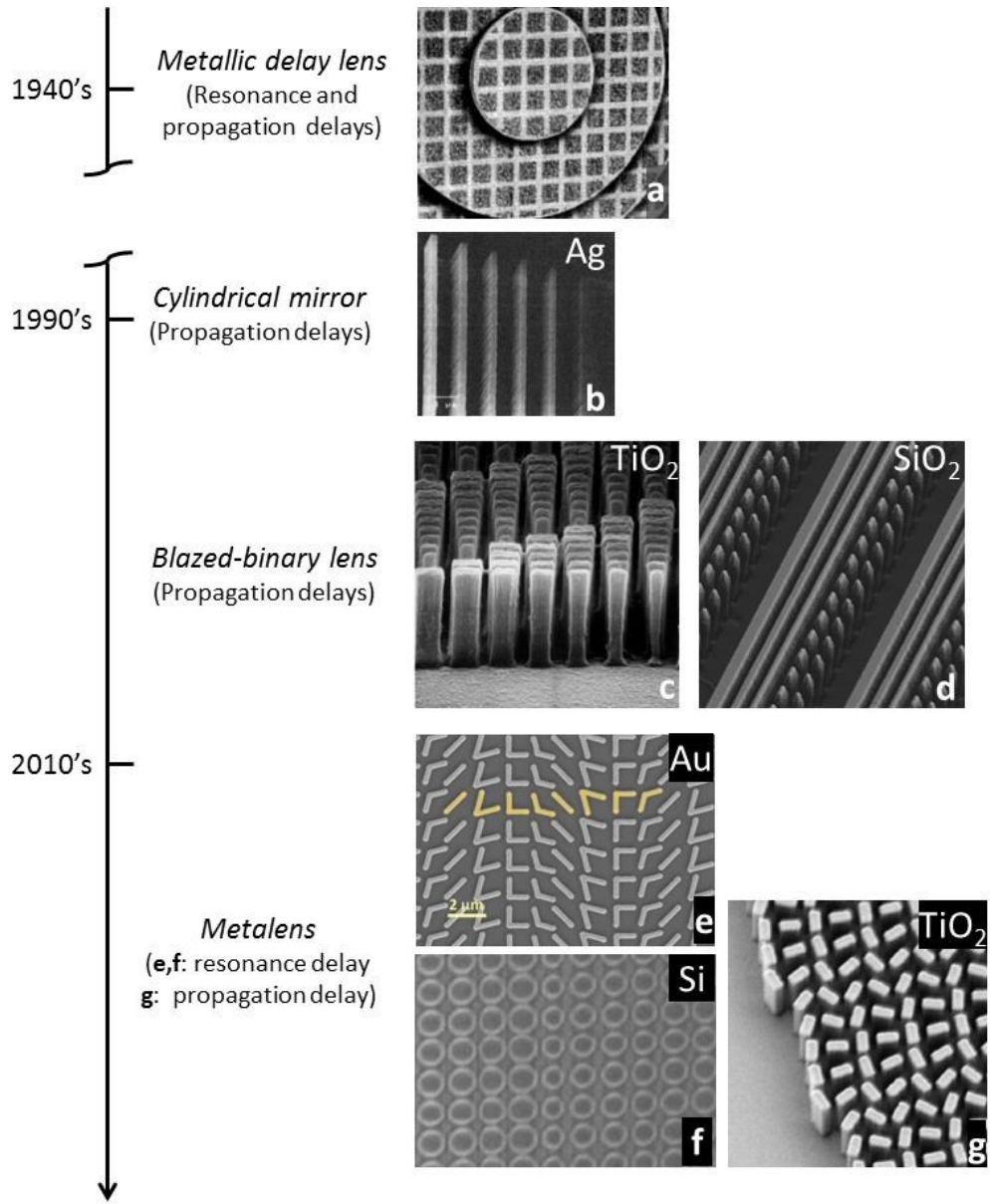

| Reference | Efficiency | mode | material | λ | polarization | $t/\lambda$ |
|---|---|---|---|---|---|---|
| **a:** [Koc48] | not relevant | T | composite | 7 cm | insensitive | not relevant |
| **b:** [Kip94] | 80% | R | Ag | 10.6 μm | linear | 0.7 |
| **c:** [Lal99] | 80% | T | TiO$_2$ | 633-850 nm | insensitive | 1.15 |
| **d:** [Zei10] | >80% | T | SiO$_2$ | 850 nm | insensitive | 2.1 |
| **e:** [Yu11] | weak | T | Au | 8 μm | linear (cross) | 0.006 |
| **f:** [Yu15] | 50% | T | Si | 710 nm | insensitive | 0.18 |
| **g:** [Kho16] | 66-86% | T | TiO$_2$ | 405-660 nm | circular (cross) | 0.9 |

**Fig. 3. Historical fresco of wavefront shaping with subwavelength binary nanostructures at optical frequencies.** R and T mean reflection and transmission mode operation, respectively, and "cross" refers to incident and diffracted beams which are mutually cross-polarized.

# 4. Limitations and perspectives

## 4.1 Phase control at high spatial resolution

The art of designing a high-NA diffractive lens relies on the capability of imprinting rapidly varying phases by exploiting subwavelength guidance, or more generally confinement, in dielectric nanostructures. The issue faces two conflicting perspectives.

On one side, one would prefer to have small spacing between nanostructures (right panel in Fig. 4). Then, the phase would be finely sampled, which is important especially for achieving high efficiency. For instance, we know that the efficiency of diffractive element with a slowly-varying phase that is sampled with 4 uniformly-distributed values in the interval [0, $2\pi$] is only 81%, and that 16 values are required to reach a 99% efficiency [Swa89]. On another side, in addition to fabrication issues, there is a lower bound for the spacing. Indeed, one expects that every nanostructure be electromagnetically independent from its neighbors [La99a], so that the phase at the local control points is really implemented locally at the level of every individual nanostructure rather than over an extended area covering a few neighboring nanostructures. These requirements apply to both waveguide-type and resonant metalenses.

Indeed reaching a good balance between the two conflicting perspectives requires nanostructures with a strong transversal confinement of light, otherwise the sampling points do not independently control the phase. How to implement such confinements? The whole literature on waveguide-type or resonant metalenses has consistently carried one and the same response over the years: One should use nanostructures with a high-index-contrast ratios by etching films, e.g. titanium dioxide [Lal98,Kho16] or semiconductor [War96,Fat10,Vo14,Arb15] layers with a high-index often deposited on a substrate with a lower refractive index. That guarantees that the evanescent electromagnetic tails of every nanostructure decrease rapidly and are essentially damped before reaching the neighbors. The reason is obvious for metalenses relying on propagation delays controlled by tiny waveguiding pillars, and becomes clear for resonant metalenses by considering that a Mie-like resonance of various resonator geometries, e.g. rods, disks, pillars, or even spheres, can be seen as a Fabry-Perot resonator formed in short waveguides with arbitrary semi-nanoparticles forming the terminations, see for instance the detailed discussions conducted in [Has11,Tra15].

## 4.2 Phase encoding with monomode and multimode operation

The principles of operation of resonant and waveguide-type metalenses are contrasted. It is important to well understand differences and similarities, together with the respective advantages and limitations.

### 4.2.1 Waveguide-type metalenses

Figure 4 summarizes the main design constraints on the spacing values for waveguide-type metalenses that offer either polarization insensitive or Berry-phase vortex operation with propagation delays. In the right panel, small spacings are considered. The phase sampling is fine, but the nanopillars are coupled electromagnetically and, due to the coupling, the supermode propagation constant $k_z$ varies with the parallel wavevector $k_{//}$ of the incident plane wave. The shadow zone is not completely eliminated and the efficiency is low [Lal99]. In the central panel, the spacing is adequate: The nanopillars are uncoupled, their normalized propagation constant being equal to the effective index of the isolated waveguide. Thus the imprinted phase becomes independent of $k_{//}$ implying that the same pillar structure can be used for a broad range of incidence angles and may effectively either focus any obliquely incident plane wave, or collimate any diverging beam. The left panel corresponds to spacings that are too large. Two supermodes are propagating and the imprinted phase that depends on both modes is difficult to accurately controlled, especially if $k_{//}$ varies. Safely imprinting the wrapped phase requires a monomode propagation for light in the pillars, with a pillar spacing smaller than an upper bound, called the structural cutoff in [Lal99], which guarantees that a single supermode per polarization propagates in the pillar array, all the other supermodes being evanescent.

Overall, Fig. 4 evidences the narrow working window for suitably selecting the nanostructure spacing – an essential parameter – and faithfully imprinting the desired phase profile. The narrow window led the present authors to choose a transparent material with the highest possible index (TiO$_2$ in the visible) so as to maximize the transversal field confinement and to choose a spacing-to-wavelength ratio of 0.47 for the TiO$_2$ blazed-binary elements of Fig. 3c. We note that any attempt towards the use of larger spacings, to relax the fabrication constraints for instance, systematically led to the observation of lower efficiencies, and this since Chen's first attempt [Che96].

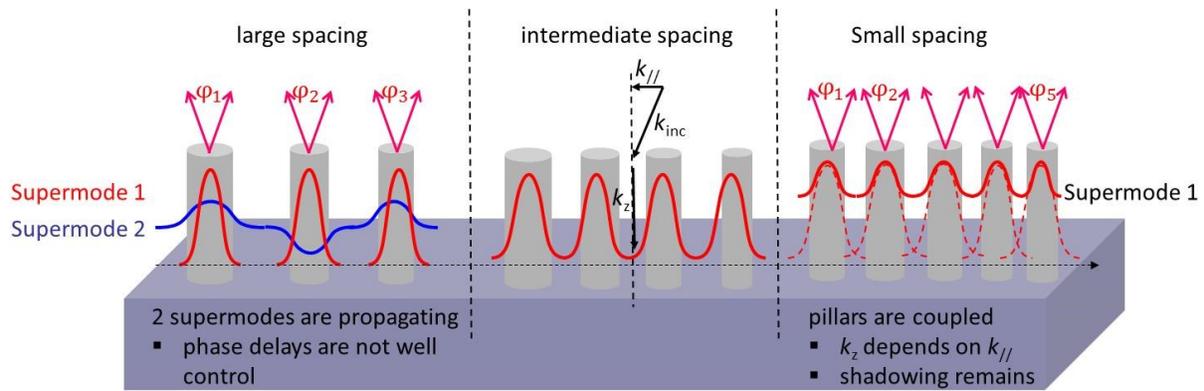

**Fig. 4. Critical choice of the nanostructure spacing in waveguide-type metalenses.** Right panel: small spacings. The phase sampling is fine but the nanopillars are coupled electromagnetically. Central panel: The nanopillars are uncoupled. Their normalized propagation constant is equal to the effective index of the isolated waveguide, and the imprinted phase is independent of $k_{//}$. Left panel: large spacings. Two supermodes are propagating and the imprinted phase that depends on both modes is not accurately controlled.

### 4.2.2 Resonant metalenses

The design recipes of resonant metalenses are different from those of waveguide-type metalenses, and are better understood in the language of localized 3D resonator modes. By combining two frequency-matched resonances, each covering a standard phase range of $\pi$, a $2\pi$-phase excursion is imprinted by controlling the nanoresonator dimensions [Vo14,De15,Yu15,Arb15,Zha16]. The precise physical mechanism that is exploited can be well understood from a simple model, assuming that the resonant metasurface is equivalent to an idealized subwavelength array of metaatoms composed of a pair of electric and magnetic in-plane dipoles with Lorentz polarizabilities [De15].

Two resonant geometries have been mainly considered so far. Microposts offer high efficiencies. However, they can be used only for a reduced range of duty cycles to avoid spurious resonance with nearly null transmittance [Arb15,Zha16,Vo14], and the aspect ratio is only weakly relaxed in comparison with the monomode guidance approach. Nanodisks [De15,Yu15] appear more promising as they allow for significantly smaller aspect ratios that are fully compatible with standard industrial silicon technology, see Fig. 3f. Although the measured efficiencies to date are smaller than 60%, ultra high performance with nearly-100% transmission efficiency appears realistically achievable with Si nanodisks embedded into an optimized dielectric environment, as predicted by computational results [De15].

If two localized symmetric modes coexist for a given wavelength, an antisymmetric mode must also exist at the same wavelength. This assertion can be intuitively understood by considering the equivalence between Mie-like localized resonances and Fabry-Perot resonances [Tra15]. Since the two localized symmetric modes are necessarily formed by the cycling of two symmetric guided modes, we know from waveguide theory, that an antisymmetric guided modes should also exist at the same wavelength and it is thus expected that another localized mode, formed by the cycling of the antisymmetric guided mode, exist.

To evidence the existence of this third localized mode and to study its impact, we consider the state-of-the-art Huygens' metasurface proposed in [Dec15] and compute its transmittances at normal and small oblique incidences. The transmittances are displayed in Fig. 5. At normal incidence (black curve), consistently with the red curve in Fig. 5c in [Dec15], nearly 100% transmission efficiency is impressively observed over a $2\pi$-phase excursion (we have not plotted the phase-variation but the interested reader may refer to the blue curve in Fig. 5c in [Dec15]). However, the transmittance spectra computed at slightly oblique incidences (blue and dotted-red curves) display dips that are typically due to the excitation of antisymmetric resonances not excited at normal incidence. These dips, which in addition are strongly polarization dependent, considerably reduce the efficiency. A different way to understand their existence can be well grasped from the physics of resonant high-contrast gratings with symmetric and antisymmetric Bloch modes, as discussed in detail in [Lal06].

To our knowledge, the detrimental impact of antisymmetric localized modes has not been discussed in the literature of resonant metalenses [Vo14,De15,Yu15,Arb15,Zha16]. It may explain why, despite the ease of fabrication of low-aspect-ratio nanodisks, the measured efficiency, inevitably based on imperfectly collimated and slightly tilted incident beams, is significantly lower than the efficiency computed for normal incidence. Antisymmetric localized modes may also severely limit the applicability of resonant metalenses, if they are not taken into account in the design.

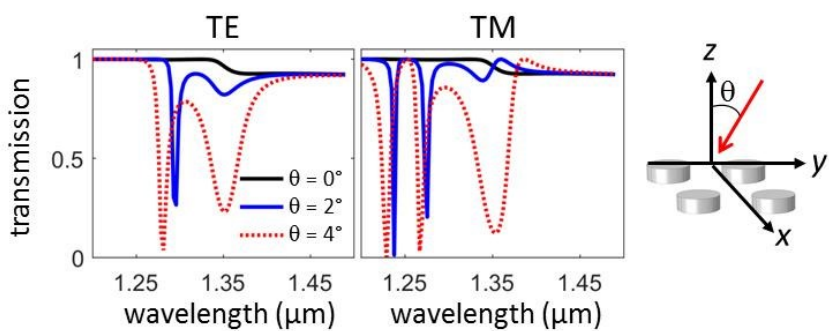

**Fig. 5. Theoretical transmittance spectrum of the Huygens' metasurface composed of a 2D array of Si-disks.** The geometrical and material data are taken from [Dec15]: the disk diameter is 484 nm, the disk thickness is 220 nm, the *x*-and *y*-periods are 668 nm, and the 2D array is assumed to be buried in material with a refractive index 1.66. The transmission efficiency for normal incidence (black curve) is nearly 100%, in quantitative agreement with the computational results shown in Fig. 5c in [Dec15]. The blue curve and red dotted curves (not computed in [Dec15]) are obtained for slightly oblique incidences. TE and TM polarizations refer to an incident plane wave (red arrow

in the inset) with an electric field parallel and perpendicular to the (*z*,*y*) plane. The computations are performed with a Fourier Modal Method [Pop00].

### 4.3 High-NA high-efficiency metalenses

#### 4.3.1 Efficiency

Large efficiencies, defined as the ratio between the power funneled into a specific desired order and the incident energy impinging on the diffractive element, are required for most applications. It is this efficiency that was considered in the previous figures and tables. It includes reflection losses. For imaging applications, it is acceptable that some power is lost by reflection/absorption, and another figure of merit, the relative efficiency defined as the fraction of the transmitted power that is effectively diffracted into the desired diffraction order, may be the proper parameter to maximize. Indeed, light that is not scattered into the desired diffraction orders feeds spurious diffractions orders, a set of diverging or converging spherical waves with focal lengths that are integer sub-multiples of the nominal focal length. They deteriorate image quality by effectively superimposing a coherent background.

#### 4.3.2 State of the art

In the perspective of accessing the present status of the field of high-NA high-efficiency metalenses, it is appropriate to analyze the recent flagship publication [Kho16] reporting metalenses respectively designed for operation at blue, green and red wavelengths with a huge *NA* of 0.8. Table 1 puts into perspective the achievements in [Kho16] in comparison to previous state-of-the-art results [Lal99]. One should note here that [Arb15] has reported an even higher *NA* (0.88) in a polarization insensitive metalens used in the telecommunication infrared range.

Except for the phase encoding approaches, the works in [Kho16] and [Lal99] present many similarities, since both considers waveguide-type metalenses, nanostructures formed with $TiO_2$ pillars with similar aspect ratios and almost identical spacing-to-wavelength ratios especially for the "blue" lens. There is no apparent real conceptual rupture. However, with a 86% efficiency, the blue lens is quite remarkable. It would be interesting to know what is the effective efficiency, but it could be reasonably expected that at least 90% of the transmitted power is effectively diffracted into the desired diffraction order, a remarkable value in view of the high NA that is achieved.

It is interesting to understand why the blue metalens has a slightly better diffraction efficiency than the blazed-binary diffractive elements. One possibility is a better fabrication technology. Another hypothesis is that polarization independence requires that the phase is varied by changing the size of the pillars. In total not all pillars are really independent from each other, especially tiny ones. In contrast, all pillars of the metalens are identical up to a rotation, and independence might well be better satisfied.

Also the 86% efficiency represents an averaged value over the entire lens, and it would be interesting to appreciate what is the actual efficiency achieved at the outer zones of the lenses, which indeed is not easy. The focal spot FWHM, as reported in [Kho16], is a good indicator of diffraction limited operation only when the focal spot matches an Airy pattern. But the focal spot sidelobes are enhanced compared to an Airy pattern, an observation that is consistent with the observation of a departure of the MTF (Fig. 2 of [Kho16]) from that of a diffraction limited lens. In other words, the characterization of metalenses is an emerging problem that [Kho16] has started addressing, but that will require more work in the metalens community.

| | Blazed binary elements [Lal99] | Metalenses [Kho16] | | |
|---|---|---|---|---|
| | 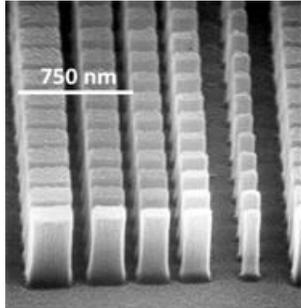 | 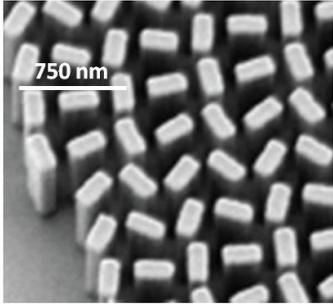 | | |
| Material | $TiO_2$ | $TiO_2$ | | |
| Wavelength (nm) | 633 – 860 | 405 (blue) | 532 (green) | 660 (red) |
| Spacing $S/\lambda$ | 0.47 | 0.49 | 0.61 | 0.65 |
| Efficiency | 80% (absolute) | 86% | 73% | 66% |
| Polarization | Insensitive | Circular (cross) | | |
| NA | 0.64 | 0.8 | | |

**Table 1.** Comparison between the blazed-binary gratings and lens manufactured in [Lal99] and the 2016 metalenses in [Kho16]. The phase encoding technique used for the metalens operates with circularly polarized light, whereas the technique in [Lal99] is polarization insensitive. Absolute efficiencies are considered in [Lal99], whereas in [Kho16] the efficiency relates a beam that has already experienced a ~4% reflection loss.

Achieving high efficiency at high NA or deviation is really challenging. The width *w* of the outer zones of a lens is actually 1/NA when normalizing to the wavelength, implying that for NA=0.8 only 2.5 (resp. 2.0) pillars on average are used to sample the phase of the outer zone of the blue (resp. green) metalenses. In [Lal99,Lee02], substantial computations and optimizations were performed to evaluate what could be the best diffraction efficiencies achievable with high-NA blazed-binary lenses fabricated with $TiO_2$ pillars by studying gratings with the same period as the zone widths. Essentially, it was found that the relative efficiency slowly decreases from 100% for low *NA* to 85% for $NA = 0.5$. For larger *NA*'s, the zone width becomes comparable to the wavelength and resonance effects occur, producing chaotic unintuitive variations of the efficiency. Nevertheless, it was possible to design gratings with a deviation angle $\theta = 51°$ ($NA = 0.78$) with a relative efficiency of 83% and to demonstrate experimentally a blazed-binary grating with a deviation $\theta = 40°$ ($NA = 0.64$) and an efficiency of 80%. Further simulations seem to indicate that, due to fabrication errors of the precise dimensions of the nanopillars and residual roughness, the measured efficiencies are 5-7% smaller than the theoretical values and ≈ 5% smaller than the relative efficiencies.

These results suggest that it would be very instructive to redesign a series of gratings with varying periods for the technology used in [Kho16], to fabricate and measure their efficiencies in order to quantitatively predict the efficiencies of the outer zones of metalenses fabricated with the technology in [Kho16]. That is essential to in-fine predict the image quality of high-NA metalenses.

### 4.4 Field of view

Commercial lenses for imaging applications are composed of several elements made from diverse materials, and some of their surfaces exhibit non-spherical shapes. Spatial separation of the elements along the lens axis is required to correct for aberrations when imaging at wide angles and large fields with diffraction-limited focusing spots. In sharp contrast, all metalenses fabricated up to now using the various approaches mentioned above use a single element to imprint the phase. Whichever the method used for encoding the phase, that precludes high quality imaging over a large angular extent, as we shall illustrate now.

The issue can be illustrated by considering Fig. 6, which shows focusing at normal incidence (dashed blue) and at oblique incidence (solid red). Denoting by $x$ and $y$ the pupil coordinates, elementary geometrical calculations show that the phase-difference between points A and B that should be implemented for perfect focusing at normal incidence is $(\varphi_A - \varphi_B) = k_0(f^2 + x^2 + y^2)^{1/2} - k_0 f$, whereas it should be $k_0\left(f^2 + (x - f\tan(\theta))^2 + y^2\right)^{1/2} - x\sin(\theta)$ at oblique incidence $\theta$. Here, B is an arbitrary point in the pupil, that can move up to the border as shown in the figure. The departure between these two expressions can easily be expanded, for a small pupil size $r$ and a small angle $\theta$, into the standard aberrations of coma, field curvature, and astigmatism. As said, those are normally compensated in a compound lens by taking benefit of the distance between the lens elements. In a single lens, those aberrations increase quite rapidly with increasing aperture or field angle, and there is not much that can be done about it.

Therefore, for a large NA, the field of view is very small. For example, for NA = 0.5, with a focal length of 1 mm, the low aberration field of view with a path delay error not exceeding one wavelength, in the visible, is restricted to less than 5 µm. This implies that single-interface lenses might be useful for point-to-point interconnects but not for high-performance imaging. Having that consideration in mind, it is inappropriate and even misleading to compare the imaging performance of a single lens and multi-lens objectives corrected for aberrations, as done in [Kho16].

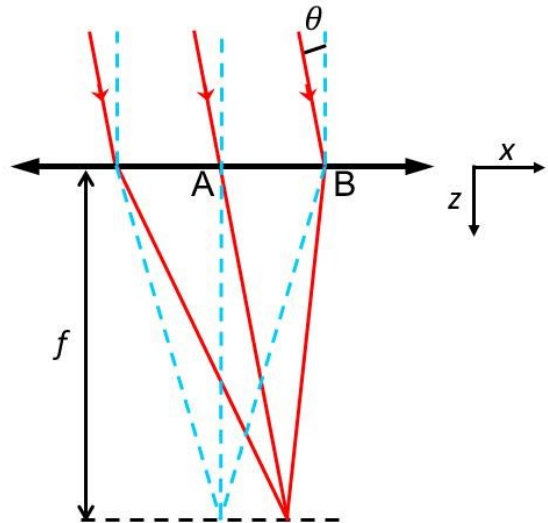

**Fig. 6. Geometrical aberrations with single-surface imaging.**

### 4.5 Broadband operation

The $2\pi$-phase excursion is in general achieved at a single wavelength, the blaze wavelength used for the design. Illuminated at another wavelength, the $2\pi$-phase excursion is not maintained and light is no longer funneled into a single order; the deviation results in the apparition of deterministic scattering into spurious diffraction orders, which represents a severe limitation for optical imaging systems designed to operate over finite spectral bands. Because the dispersion of most materials is weak away from their absorption bands, the efficiency dependence with the wavelength obeys a universal law $\eta = \text{sinc}^2(1 - \lambda_0/\lambda)$, $\lambda_0$ being the blazed wavelength for which the $2\pi$-phase excursion is achieved [Swa89].

Metalenses bring a new degree of freedom. Because the artificial dielectric is much more dispersive than a bulk material, one should consider the dependence of the effective index with the wavelength, and the optical phase difference is now written

$$\frac{2\pi}{\lambda}\Delta = \frac{2\pi}{\lambda}(n(\lambda) - 1)t. \qquad [3]$$

With this extra degree of freedom, it is possible to exploit the highly chromatic behavior of semiconductor structures with dimensions only slightly smaller than the wavelength to make the product $\frac{2\pi}{\lambda}(n(\lambda) - 1)$ nearly independent of wavelength. The idea initiated in the context of broadband subwavelength phase-plate gratings [Kik97] was tested in [Rib13] for metalenses composed of a careful combination of micropillars and microholes. The device has been characterized in the thermal infrared (band III) and blazing over nearly one octave was observed experimentally.

Not only does the efficiency of a diffractive lens varies with the wavelength, but so does its focal length, $f(\lambda_2)\lambda_2 = f(\lambda_1)\lambda_1$. Actually, the focal length decreases with wavelength, in sharp contrast with refractive lenses based on various glasses for which the refractive index (resp. focal length) decreases (resp. increases) with wavelength. It is this complementarity that is exploited in achromatic hybrid doublets to reduce the weight and size of optical systems [Blo95].

Instead of using the complementary strengths of refractive and diffractive lenses in hybrids, in the spirit of completely replacing refractive lenses by diffractive ones, one may envision to manufacture lenses with focal lengths that are independent of the wavelength [Kho15]. That cannot be performed over a continuum of wavelengths, but it can be envisioned for a discrete set of wavelengths, at blue, green and red wavelengths (RGB) for instance. In Fig. 7, we sketched the pressure placed upon the designer to imagine a multiplexed metalens that imprints three graded phases independently at three different wavelengths. Starting from a classical graded metalens in (a), each nanostructure should be replaced by three even-smaller colored nanostructures with effective cross sections that are more than 10 times larger than their physical cross sections, so that all the light incident on every unit cell (dashed-line box in (a) or (b)) is funneled into the nanostructure to imprint the desired phase at the nominal wavelength. In the spectral domain, see (c), each colored nanostructure should be wavelength-selective to avoid inevitable spectral crosstalk due to the spurious scattering by neighbor nanostructures designed for operation at a given wavelength and illuminated at another. All in all, fulfilling that dual requirement for every nanostructure represents a real challenge with the design of deep subwavelength nanoresonators that efficiently funnel light in the spatial domain and provide high Q resonance in the spectral domain. Even if the requirements are physically linked by space-bandwidth products arguments, high efficiency and faithful operation will be challenging to reach.

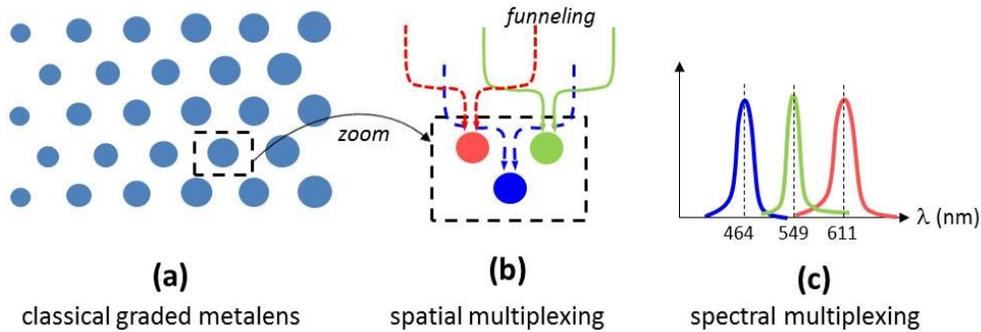

**Fig. 7. Big design challenge for metalens operating independently at three different wavelengths**. To avoid crosstalk, the multiplexing should fulfill drastic requirements in space (antenna funneling) and spectral (high Q) domains.

## 5. Conclusion

The previous analysis leads us to the following conclusive points:

- Focusing light with high-efficiency and high-*NA* is feasible at one wavelength in a point-to-point interconnection scheme with a single interface patterned with subwavelength high-index structures that guide light. This property, which cannot be achieved with classical échelette diffractive optical elements, has been known since the late 90's [Lal99, La99a].

- High performance imaging (with high-*NA*) is not possible with a single diffractive (or refractive) interface, see Section 4.3. Manufacturing high-efficiency and high-*NA* metalenses for high-performance imaging requires to correct geometrical aberrations with several diffractive surfaces, as recently demonstrated in [Arb16]. It would be interesting to know what could be the minimum separation distance between the surfaces, but thinness of the global optical system and high performance imaging over a wide field of view seem to be two incompatible requirements [Gis16].

- Broadband blazing (typically over one octave) with high relative efficiency has been demonstrated for applications in the thermal infrared with metasurfaces that exploit the highly chromatic behavior of subwavelength structures, see Section 4.4. Unfortunately the demonstration uses holes, instead of pillars, in the shadow zone and since holes do not guide light efficiently, high efficiency is achieved only for devices with low deviation angles $\theta$. At this stage, there is no known metasurface design that offers broadband blazing and high *NA*. Note that multilayer diffractive échelette-type components exist for broadband operation with ultra-high efficiency (>95%) and low-*NA* [Nak02]. Such components have been pioneered by Cannon in the EF 400mm f/4 DO IS USM teleobjective lens.

- Manufacturing cost matters. Echelette diffractive optical elements are presently manufactured at very low-cost in plastics essentially, with replication techniques such as roll-on, embossing ... [Gal97]. There is a common underlying cost to manufacturing metalenses. The latter require subwavelength features with large aspect ratios, see the last column in the Table of Fig. 3, making mass production a real challenge. In this line of thought, fabrication constraints appear relaxed with metalenses implementing $2\pi$-phase excursion with a combination of two resonances rather than with a propagation delay. The fabrication cost of metalenses based on plasmonic nanoantennas that may potentially be patterned on a low-cost flexible substrate with roll-on techniques could be markedly lower. However, in addition to efficiency issues, the low defect requirement needed to lower inevitable scattering due to roughness and impurities seems challenging as well.

Aside from those concerns on imaging issues with metalenses, niche applications may exist. Metasurfaces that provide fancy polarization behaviours with birefringent nanopillars, which cannot be easily implemented with classical refractive or échelette components, may offer new opportunities as recently demonstrated for single molecule spectroscopy for instance [Bac16].

**Acknowledgements**

The authors have appreciated the repeated, fruitful and friendly discussions with Jacob Khurgin, Yuri Kivshar, Erez Hasman, Sergei Bozhevolnyi, Vlad Shalaev, Arka Majumdar, Thomas Krauss, Pavel Cheben, Andrei Faraon, Arseniy Kuznetsov, who helped much refining our opinion on this booming field. They also wish to thank five anonymous reviewers for their kind interaction, repeated questioning and persuasive comments.